\documentclass[twocolumn,aps,floatfix,pra]{revtex4}
\usepackage{amssymb,amsfonts,mathrsfs}
\usepackage{graphicx,bm}
\usepackage[T1]{fontenc}

\begin{document}
%draft

\title{Bose statistics and classical fields}
\author{Emilia Witkowska$\,^1$, Mariusz Gajda$\,^{1,}\,^2$ and Kazimierz Rz\k{a}\.zewski$\,^{2,}\,^3$}
\affiliation{
\mbox{$^1$ Institute of Physics, Polish Academy of Sciences, Aleja Lotnik\'ow $32/46$, 02-668 Warsaw, Poland}
\mbox{$^2$ Faculty of Mathematics and Sciences, Cardinal Stefan Wyszy\'nski University, Warsaw, Poland}
\mbox{$^3$ Center for Theoretical Physics, Polish Academy of Sciences, Aleja Lotnik\'ow $32/46$, 02-668 Warsaw, Poland}}
\date{\today}

\begin{abstract}
Classical fields counterpart of the ideal Bose gas statistics in a trap is investigated by performing calculations in the canonical ensemble. There exists the optimal cut-off which allows to match the full probability distribution of the condensate population by its classical counterpart. Universal scaling of that cut-off with temperature and dimensionality is derived.
\end{abstract}
\pacs{
% BEC
03.75,
%Quantum statistical mechanics
05.30.}
\maketitle

At the end of XIX century equilibrium thermodynamics of the electromagnetic field, the theory of a so called black body radiation, suffered for the persistent ultraviolet divergence. It was Max Planck \cite{Planck} who cured the problem introducing a concept of light quanta later called photons. With the help of photons the energy carried by high frequency degrees of freedom was tamed.

More than 20 years later Einstein \cite{Einstein}, stimulated by preliminary ideas of Bose \cite{Bose} introduced statistical properties of the ideal gas of indistinguishable massive particles that are now called bosons. In particular it was shown that entirely due to the indistinguishability, when cooled, bosons tend to assemble at the ground  state of the binding potential at relatively high temperature.

This phenomena called Bose-Einstein condensation was finally realized in the seminal experiments in 1995 \cite{pierwszy kondensat}. Of course realistic theory had to depart from the ideal gas and was to account for interactions. While a well known Gross-Pitaevskii equation combined with small excitations - the Bogoliubov approximation \cite{Bogolibow} - forms a good basis of analysis near absolute zero temperature, many experiments require a method valid for higher temperatures. Several groups proposed a classical field approximation as an effective approach to nonzero temperature Bose gas \cite{pola klasycze}. This is like traveling the road from Jaynes to Max Planck backward in time -ignoring the granular character of atoms and stipulating that amount of atoms may be changing in a continuous way. Understandably equipartition of the energy between the modes leads to the ultraviolet divergence just as it was for classical radiation. Hence, the authors introduce a short wave length cut-off to cure the problem. Different authors and some of them differently in different papers choose the value of the cut-off \cite{{witkowska zawitkowski}, {pola klasycze}}.

It is the purpose of this Rapid Communication to show, based on exactly soluble models of the ideal gas, that there exists the optimal cut-off that allows to match the full probability distribution of the number of condensed atoms by its classical fields counterpart. We give a scaling of this cut-off with temperature and dimensionality of the binding potential. Of particular interest is the population of the last state retained in the optimal classical fields approximation. It depends on dimensionality but not on temperature.

\begin{figure}[!]
\centering
{\includegraphics[width=0.45\textwidth, height=0.45\textheight, keepaspectratio]{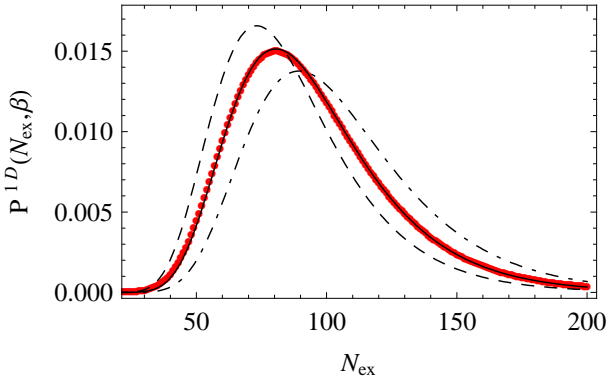}}
\caption{(Color online) Probability distribution of having $N_{ex}$ bosons outside the ground state of a 1D trap. Points represent the exact quantum distribution while lines are its classical counterparts with $\hbar \omega \beta K_{max}=0.9$ -- dot-dashed line, $\hbar \omega \beta K_{max}=1$ -- solid line and $\hbar \omega \beta K_{max}=1.1$ -- dashed line. Total number of bosons is $N=1000$ and at $\hbar \omega \beta =0.04$.}
\label{fig: distribution}
\end{figure}

We start our analysis with the one-dimensional (1D) Bose gas in a harmonic trap of energy $E=\sum_{j} n_j \epsilon_j$, where single-particle energies $\epsilon_j=\hbar \omega j$ \cite{energy note}. In order to obtain the probability distribution we first consider partition functions. The canonical ensemble partition function for a system with $N_{ex}$ excited bosons and temperature $T$ of 1D harmonic oscillator is \cite{Holthaus Weiss}:
\begin{equation}
Z_{ex}(N_{ex},\beta)=\sum_{n_1=0}^{\infty}\sum_{n_2=0}^{\infty}\dots e^{-\beta E} \delta_{\overline{N}_{ex}, \, N_{ex}} \,
\label{eq: ZKW}
\end{equation}
where $n_j$ are populations of $j$th state, $\overline{N}_{ex} = \sum_{j=1}^{\infty}n_j$ is a number of bosons and $\beta=1/k_B T$. Kronecker delta function in (\ref{eq: ZKW}) enforces the condition of $N_{ex}$ thermal bosons. In terms of these functions probability distribution of having $N_{ex}$ thermal bosons is $P^{1D}(N_{ex},\beta)=Z_{ex}(N_{ex},\beta)/Z_N$ with $Z_N=\sum_{Nex=0}^{N}Z_{ex}(N_{ex},\beta)$. Introducing an integral representation of the delta function $\delta_{a, \, b}= \frac{1}{2 \pi} \int_{0}^{2 \pi} e^{i x(a-b)} \mathrm{d} x$ and calculating the integral we obtain:
\begin{equation}
P^{1D}(N_{ex},\beta)=\frac{1}{Z_N} \sum_{j=1}^{\infty} e^{-\beta j N_{ex}} \prod_{l \ne j}^{\infty} \frac{1}{1-e^{-\beta(\epsilon_l-\epsilon_j)}} \, .
\end{equation}
Similar probability distribution can be calculated for the classical counterpart. In this case the canonical ensemble partition function must take into account continuous values of populations while energy levels are quantized as previously. For the classical counterparts sums are replaced by integral and the canonical ensemble partition function is:
\begin{equation}
\mathcal{Z}_{ex}(N_{ex},\beta) = 
\int \frac{\mathrm{d}^2 \alpha}{\pi} \int \frac{\mathrm{d}^2 \alpha}{\pi} \dots e^{-\beta \mathcal{E}} \delta(\overline{\mathcal{N}}_{ex}-N_{ex}) \, ,
\label{eq: ZKL}
\end{equation}
where amplitudes $|\alpha_j|^2 \in (0, \infty)$, the energy $\mathcal{E}=\sum_{j=1}^{K_{max}} |\alpha_j|^2 \epsilon_j$ and $\overline{\mathcal{N}}_{ex}=\sum_{j=1}^{K_{max}}|\alpha_j|^2$. Notice, that we include a finite numbers of states up to $K_{max}$ only. Populations of the higher energy states are set to zero and not taken into account within the classical approximation. The integral representation of the Dirac delta function $\delta(a-b)=\frac{1}{2 \pi}\int_{-\infty}^{\infty} \mathrm{d} x e^{i x (a-b)}$ simplifies calculations which give:
\begin{equation}
\mathcal{P}^{1D}(N_{ex},\beta)=\frac{1}{\mathcal{Z}_N} 
\sum_{j=1}^{K_{max}} e^{-\beta j N_{ex}} \prod_{l \ne j}^{K_{max}} \frac{1}{\beta(\epsilon_l-\epsilon_j)} \, .
\label{eq: Pclassical}
\end{equation}
Note, that this formula leads to the equipartition of energy in every single-particle mode, i.e. $\langle n_j \rangle \epsilon_j =  k_B T$, where $\langle n_j \rangle$ is the mean occupation of the state of energy $\epsilon_j$. This feature is characteristic for thermal state of a classical system. Probability distributions $P^{1D}(N_{ex},\beta)$ and its classical counterpart $\mathcal{P}^{1D}(N_{ex},\beta)$ are shown in Fig.\ref{fig: distribution}. One can see that there exists an optimal value of the cut-off which in this case is equal to $K_{max}=1/\hbar \omega \beta$. For this optimal value quantum distribution and its classical counterpart are almost indistinguishable. Notice, that other values of $K_{max}$ lead to a shift of the entire distribution while its shape does not change significantly.

It is interesting to analyze the system in 2 and 3 spatial dimensions. Now, the problem is more complex since we have to deal with a degeneracy. There exists excellent approximate formulas in the literature and we find the result of \cite{Scully} the most useful for our purposes:
\begin{equation}
P(N_{ex},\beta)=\frac{1}{Z_N}\frac{(N_{ex}+\mathcal{H}/\eta-1)! }{(\mathcal{H}/\eta-1)! (N_{ex})!} \left( \frac{\eta}{\eta +1}\right)^{N_{ex}} \, ,
\label{eq: sully kw}
\end{equation}
with parameters:
\begin{eqnarray}
\mathcal{H}(\beta)&=&  \sum_{{\bf j} \ne 0}^{\infty} \frac{1}{e^{\beta \epsilon_{\bf{j}}}-1}\, ,\label{eq: H kw}\\
\eta \mathcal{H} &=& \sum_{{\bf j} \ne 0}^{\infty} \frac{1}{(e^{\beta \epsilon_{\bf{j}}}-1)^2}\, .
\end{eqnarray}
Please note again that each term in the sum (\ref{eq: H kw}) is a mean occupation of the single-particle level ${\bf j}$  which in the classical version (\ref{eq: H kl}) implied equipartition $\langle n_{\bf j} \rangle \epsilon_{\bf j} =  k_B T$. The simple expression (\ref{eq: sully kw}) is valid for any trapping potential which controls functions $\mathcal{H}$ and $\eta$ through the single particle energy spectrum $\epsilon_{\bf{j}}$ of a trap.  Within the classical fields formula (\ref{eq: sully kw}) is still valid but parameters $\mathcal{H}$ and $\eta$ are different:
\begin{eqnarray}
\mathcal{H}_{cl}(\beta, K_{max})&=&  \sum_{{\bf j} \ne 0}^{K_{max}} \frac{1}{\beta \epsilon_{\bf{j}}}\, ,\label{eq: H kl}\\
\eta_{cl} \mathcal{H}_{cl} &=& \sum_{{\bf j} \ne 0}^{K_{max}} \frac{1}{(\beta \epsilon_{\bf{j}})^2}\, .
\end{eqnarray}
This classical version of Eq.(\ref{eq: sully kw}) is not obvious. We checked the classical counterpart of (\ref{eq: sully kw}) by comparing with (\ref{eq: Pclassical}) in 1D, and with the saddle points method \cite{saddle point method} results in 3D. For the systems with small number of bosons, as considered numerically here, the maxima of distributions coincide but there are some deviations in its width. The agreement improves for larger systems. 

\begin{figure}[!]
\centering
{\includegraphics[width=0.45\textwidth, height=0.45\textheight, keepaspectratio]{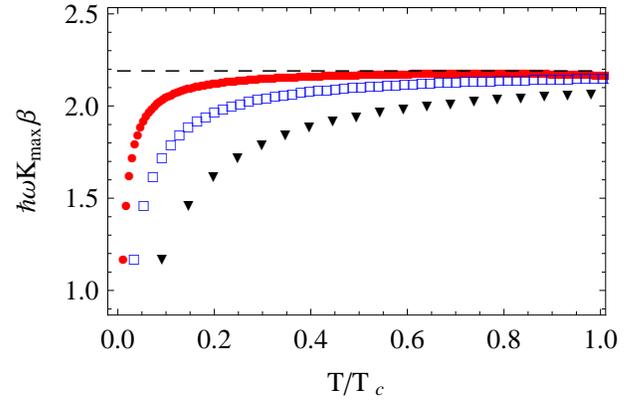}}
\caption{(Color online) Solution of the equation (\ref{eq: na Kmax}) as a function of $T/T_c$ for 3D harmonic trap and total number of bosons $N=10^3$ ($\blacktriangledown$), $N=2\, \cdot 10^4$ ($\Box$), $N=6\,\cdot 10^5$ ($\bullet$). Dashed line represents a value of $\hbar \omega K_{max} \beta$ given by the expression (\ref{eq: solution}).}
\label{fig: solution}
\end{figure}

\begin{figure}[!]
\centering
{\includegraphics[width=0.45\textwidth, height=0.45\textheight, keepaspectratio]{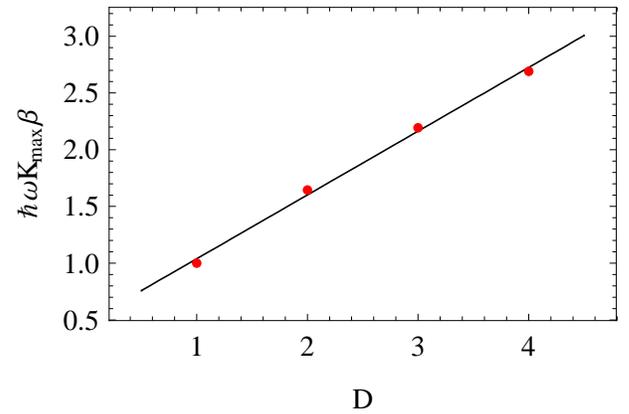}}
\caption{(Color online) Scaling of the cut-off parameter $\hbar \omega K_{max}\beta$ with dimensionality $D$. Points are given by formula (\ref{eq: solution}), line is the best linear fit: $\hbar \omega K_{max} \beta = 0.477 + 0.562 D$.}
\label{fig: Kmax linear}
\end{figure}

\begin{figure}[!]
\centering
{\includegraphics[width=0.45\textwidth, height=0.45\textheight, keepaspectratio]{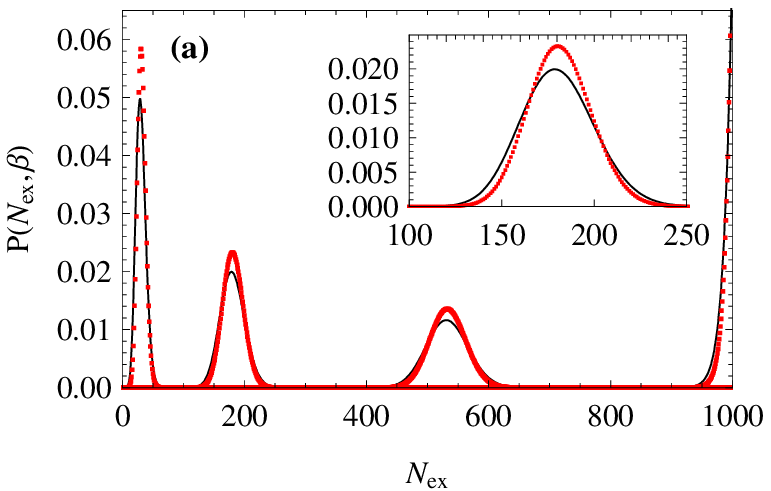}}
{\includegraphics[width=0.45\textwidth, height=0.45\textheight, keepaspectratio]{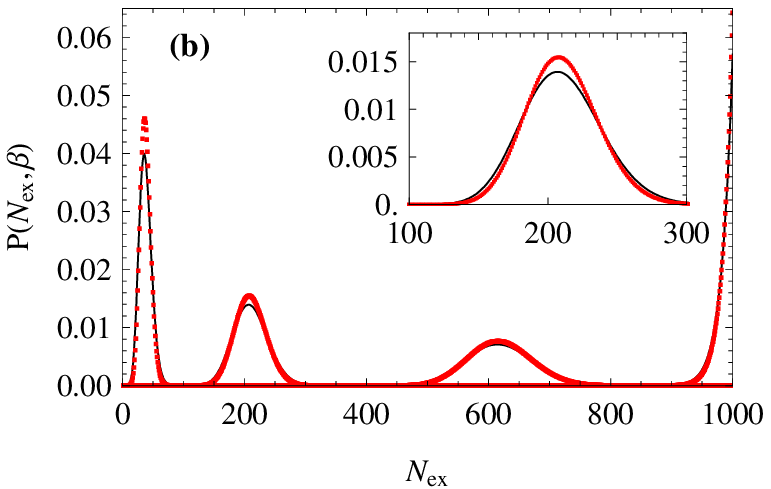}}
\caption{(Color online) Probability distribution of having $N_{ex}$ bosons outside the ground state of a 3D trap $(a)$ and 3D box $(b)$ potentials. Points are exact distribution while solid lines represent its optimal classical counterparts. Total number of bosons is $N=1000$. Notice, for 3D box potential probability distributions are wider as compared to 3D harmonic trap.}
\label{fig: distribution 3D}
\end{figure}

In order to compare the quantum and classical counterpart of the probability distribution (\ref{eq: sully kw}) we choose the cut-off $K_{max}$ in such a way that the maxima of both distributions coincide. The maximum of probability distribution (\ref{eq: sully kw}) is equal to $\mathcal{H}$ and equation for optimal $K_{max}$ leads:
\begin{equation}
\mathcal{H}(\beta)=\mathcal{H}_{cl}(\beta, K_{max})\, .
\label{eq: na Kmax}
\end{equation}
In Fig.\ref{fig: solution} the solution of the above equation is plotted for 3D geometry and different total number of bosons. For large systems the value of $\hbar \omega K_{max}\beta$ practically does not depend on $T$. Some deviations from this constant value can be observed for very low relative temperatures $T/T_c$ only ($T_c$ is a critical temperature) and this region decreases with increasing $N$. The value of $\hbar \omega K_{max} \beta$ saturates at:
\begin{equation}
(\hbar \omega K_{max} \beta)^{D-1}=\left\{
\begin{array}{ll}
1\, , & \textrm{for $D=1$,} \\
\zeta(D) (D-1) (D-1)!\, , & \textrm{for $D \ge 2$,}
\end{array}
\right.
\label{eq: solution}
\end{equation}
where $\zeta(D)$ is the Riemann Zeta function \cite{Kmax formula}. This condition can be used as a criterion for the optimal choice of the maximal energy of classical states $\hbar \omega K_{max}$. The condition (\ref{eq: solution}) defines the universal value of the cut-off $K_{max}$ which for large systems, as considered in the classical fields method, depends only on temperature and space dimensionality. In Fig.\ref{fig: distribution 3D} we compare 3D quantum probability distribution given by (\ref{eq: sully kw}) with its classical counterpart where the $K_{max}$ was chosen according to (\ref{eq: na Kmax}). We see that not only the maxima coincide but also the entire distributions are very similar. Therefore, by the appropriate choice of one parameter only, the cut-off energy of classical modes $\hbar \omega K_{max}$, we can accurately mimic the true quantum statistics by the statistics of the classical fields satisfying the equipartition of energy. This optimal cut-off scales nearly linearly with dimensionality $D$, see Fig.\ref{fig: Kmax linear}.

The same analysis can be applied for the uniform system with periodic boundary conditions, i.e. cubic box of length $L$. The single-particle energy spectrum is then $\epsilon_{\bf k}=\hbar^2 {\bf k}^2/2 m$, where ${\bf k}=2 \pi(n_{x},n_{y},n_{z})/L$ and $n_i=0,\pm 1,\pm 2, \dots$. In this case, the cut-off parameter is determined by the maximal wave-vector ${\bf K}_{max}=2 \pi (n_{max},n_{max},n_{max})/L$ and the condition (\ref{eq: na Kmax}) defines its value.  For large systems $\beta \hbar^2 {\bf K}_{max}^2/2m $ very weakly depends on temperature similarly as for the harmonic potential. Its value saturates at $\beta \hbar^2 {\bf K}_{max}^2/2m = \pi(\zeta(3/2)/4)^2 \simeq 1.34, \,  0.68, \,  0.29$ in 3D, 2D, 1D respectively what was found numerically.  In Fig. \ref{fig: distribution 3D} the quantum probability distribution and its classical counterpart are shown for the optimal cut-off given by Eq.(\ref{eq: na Kmax}).  One can see that not only the maximum but also a shape of both distributions are very close.

\begin{figure}[t]
\centering
{\includegraphics[width=0.4\textwidth, height=0.4\textheight, keepaspectratio]{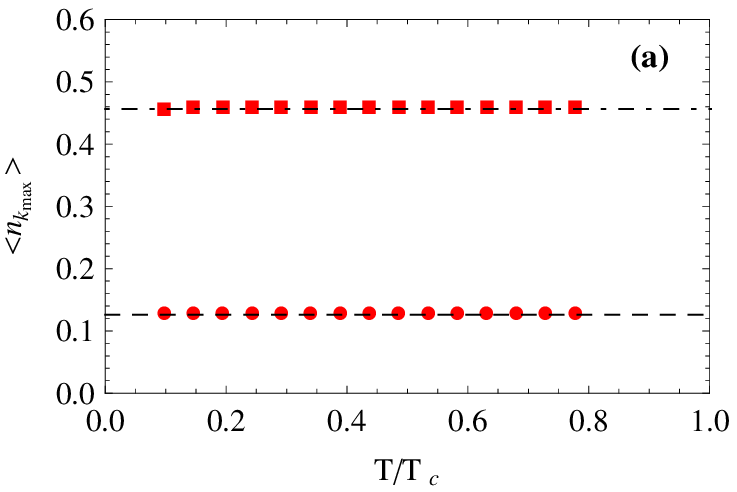}}
{\includegraphics[width=0.4\textwidth, height=0.4\textheight, keepaspectratio]{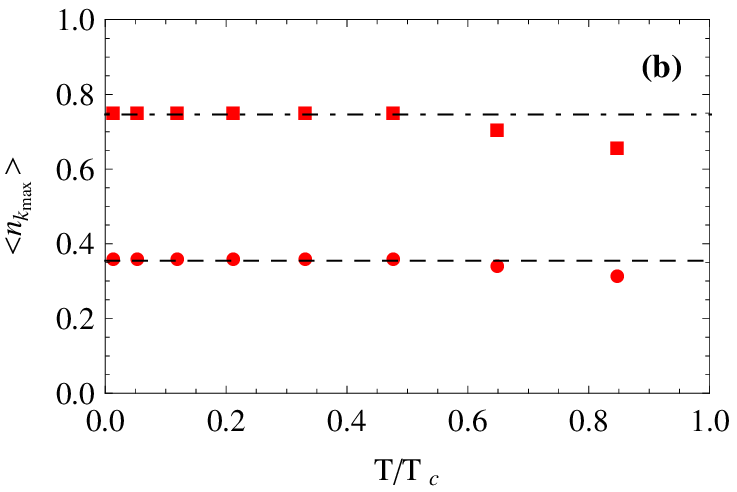}}
\caption{(Color online) Average populations $\langle n_{\bf K_{max}} \rangle$ of the final mode $K_{max}$ Eq.(\ref{eq: nkmaxkw}) ($\bullet$), and for the classical distribution Eq.(\ref{eq: nkmaxkl}) ($\blacksquare$), for 3D harmonic trap (a) and 3D box (b) potentials. Lines represent expected results in the thermodynamical limit. Number of atoms $N=1000$.}
\label{fig: nKmax}
\end{figure}

Having matched the probability distributions it is of interest to calculate the average population $\langle n_{K_{max}}\rangle$ of the final mode of the classical distribution. It can be obtained by considering probabilities $p(n_{\bf q} \ge l)$ of a mode ${\bf q}$ being occupied by at least $l$ bosons \cite{Idziaszek}:
\begin{equation}
p(n_{\bf q} \ge l) = \frac{1}{Z_{ex}(N_{ex})} \sum_{n_1=0}^{\infty} \dots  \sum_{n_{\bf q}\ge l}^{\infty} \dots e^{-\beta E} \delta_{\overline{N}_{ex}, \, N_{ex}} \, ,
\end{equation}
under the condition of $N_{ex}$ thermal bosons. The probability of having exactly $l$ bosons in mode ${\bf q}$ is $p(n_{\bf q}=l)=p(n_{\bf q}\ge l)-p(n_{\bf q}\ge l+1)$. Then the average population of ${\bf q}$ state, with the condition of $N_{ex}$ thermal bosons, is $\langle n_{\bf q} \rangle_{N_{ex}} = \sum_{l=1}^{\infty} l p(n_{\bf q}=l)$. Summation over all $N_{ex}$, with weight $p(N_{ex}|N)=Z_{ex}(N_{ex},\beta)/Z_N$ of having $N_{ex}$ atoms within $N$, gives the formula for the average population of the ${\bf q}$ mode:
\begin{equation}
\langle n_{\bf q} \rangle = \frac{1}{Z_N} \sum_{N_{ex}=0}^{N} \sum_{l=1}^{N_{ex}} e^{-\beta \epsilon_{\bf q} l} Z_{ex}(N_{ex} - l) \, .
\label{eq: nkmaxkw}
\end{equation}
For the classical distribution, where populations change in the continuous way, it is convenient to use a probability density instead. In this case, the probability of having at least $l$ bosons in mode ${\bf q}$ is $\tilde{p}_l=e^{-\beta l \epsilon_{\bf q}} \mathcal{Z}_{ex}(N_{ex}-l)/\mathcal{Z}_{ex}(N_{ex})$ while the probability density of having exactly $l$ bosons in mode ${\bf q}$ is $\tilde{p}(n_{\bf q}=l)=-\mathrm{d}\tilde{p}_{l}/\mathrm{d}l$. Then, the final expression for the average population of ${\bf q}$ mode is:
\begin{equation}
\langle n_{\bf q} \rangle_{cl} = \frac{1}{\mathcal{Z}_N} \int_{0}^{N}\textrm{d}N_{ex} \int_{0}^{N_{ex}}\textrm{d}l \,
e^{-\beta \epsilon_{\bf q} l} \mathcal{Z}_{ex}(N_{ex} - l) \, .
\label{eq: nkmaxkl}
\end{equation}
Eq.(\ref{eq: nkmaxkl}) is a continuous version of Eq.(\ref{eq: nkmaxkw}) and based on the Euler-Maclaurin summation formula the leading term of its difference is approximately equal to $\langle n_{\bf q} \rangle_{cl}-\langle n_{\bf q} \rangle \simeq 0.5$. Therefore, the average occupation of the $K_{max}$ mode within the classical fields method is larger than the quantum one. In Fig.\ref{fig: nKmax} the average populations $\langle n_{K_{max}} \rangle$ and $\langle n_{K_{max}} \rangle_{cl}$ for 3D harmonic trap (a) and 3D box potential (b) are shown as functions of the temperature $T/T_c$. Summation of populations over all excited modes must give a number of excited atoms $N_{ex}$ in both cases that is why in the classical case the populations have to be larger. The values of the average occupations are very close to expected ones in the thermodynamical limit.

In summary, we considered a classical counterpart of the probability distribution of the ideal Bose gas within the canonical ensemble. We found that there exists the optimal cut-off which allows to match the probability distribution of the condensate population by its classical fields counterparts. We found the scaling of the cut-off energies with temperature and dimensionality. This being a very simple model it nevertheless sheds some light on the optimal choice of the cut-off for weakly interacting bosons and more significantly on the very foundations of the method itself. We understood now much better that finite number of the classical fields can adequately represent the statistical properties of the full quantum many body system of bosons.

{\bf Acknowledgment}
The authors are grateful for useful discussion with M. Brewczyk. The authors acknowledge support by the Polish Government research funds for 2006-2009.


\begin{thebibliography}{99}
\bibitem{energy note}
We choose a zero of the energy scale at the single particle ground state.
\bibitem{Kmax formula}
Formula (\ref{eq: solution}) is easily derived with the help of the usual replacement of the sum by the integral in (\ref{eq: H kl}) and (\ref{eq: H kw}) and retaining the leading term in the degeneracy factor.
\end{thebibliography}
\end{document}